\documentclass{article}

\usepackage{arxiv}

\usepackage[utf8]{inputenc} 
\usepackage[T1]{fontenc}    
\usepackage{hyperref}       
\usepackage{url}            
\usepackage{booktabs}       
\usepackage{amsfonts}       
\usepackage{nicefrac}       
\usepackage{microtype}      
\usepackage{lipsum}		
\usepackage{graphicx}
\usepackage{doi}
\usepackage{CJKutf8}
\usepackage{courier}
\usepackage{float}

\usepackage{listings,jvlisting} 
\newfloat{lstfloat}{htbp}{lop}
\floatname{lstfloat}{Listing}

\title{Uncovering Black-hat SEO based fake E-commerce scam groups from their redirectors and websites}

\author{ \href{https://orcid.org/0009-0003-8447-3648}{\includegraphics[scale=0.06]{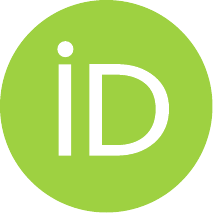}\hspace{1mm}Makoto Shimamura}\\
	Cyber Security Institute\\
	Trend Micro Incorporated, Japan\\
 	\And
    {Shingo Matsugaya}\\
	Cyber Security Institute\\
	Trend Micro Incorporated, Japan\\
        Japan Cybercrime Control center, Japan\\
 	\And
	{Keisuke Sakai}\\
	Cyber Security Control Task Force\\
	Kanagawa Prefectural Police, Japan\\
 	\And
	{Kosuke Takeshige}\\
	Community Safety Department\\
	Chiba Prefectural Police, Japan\\
    \And	
    \href{https://orcid.org/0000-0001-5596-282X}{\includegraphics[scale=0.06]{orcid.pdf}\hspace{1mm}Masaki Hashimoto}\\
	Faculty of Engineering and Design\\
	Kagawa University, Japan\\
}

\renewcommand{\shorttitle}{Uncovering Black-hat SEO based fake E-commerce scam groups}

\hypersetup{
pdftitle={A template for the arxiv style},
pdfsubject={q-bio.NC, q-bio.QM},
pdfauthor={David S.~Hippocampus, Elias D.~Striatum},
pdfkeywords={First keyword, Second keyword, More},
}

\newcommand{\numfakesite}[0]{692,865}
\newcommand{\numdomain}[0]{105,286}
\newcommand{\numemail}[0]{13,456}
\newcommand{\nummatomo}[0]{36}
\newcommand{\numid}[0]{4,958}
\newcommand{\numtotalent}[0]{123,736}

\newcommand{\numgroupsall}[0]{1,118}
\newcommand{\numthreshold}[0]{200}

\newcommand{\datadatefrom}[0]{May 20, 2022}
\newcommand{\datadateto}[0]{Dec. 31, 2024}
\newcommand{\figref}[1]{Figure~\ref{#1}}
\newcommand{\tabref}[1]{Table~\ref{#1}}

\newcommand{\nummaingroup}[0]{8}

\newcommand{\numgroupsfocused}[0]{17}
\newcommand{\numthresholdsites}[0]{2,000}

\begin{document}

\begin{CJK}{UTF8}{min}

\maketitle

\begin{abstract}
While law enforcements agencies and cybercrime researchers are working hard, fake E-commerce scam is still a big threat to Internet users. One of the major techniques to victimize users is luring them by black-hat search-engine-optimization (SEO); 
making search engines display their lure pages as if these were placed on compromised websites and then redirecting visitors to malicious sites. 
In this study, we focus on the threat actors conduct fake E-commerce scam with this strategy. Our previous study looked at the connection between some malware families used for black-hat SEO to enlighten threat actors and their infrastructures, however it shows only a limited part of the whole picture because we could not find all SEO malware samples from limited sources.
In this paper, we aim to identify and analyze threat actor groups using a large dataset of fake E-commerce sites collected by Japan Cybercrime Control Center, which we believe is of higher quality.
It includes {\numfakesite} fake EC sites gathered from redirectors over two and a half years, from {\datadatefrom} to {\datadateto}. 
We analyzed the links between these sites using Maltego, a well-known link analysis tool, and tailored programs. We also conducted time series analysis to track group changes in the groups.
According to the analysis, we estimate that {\numgroupsfocused} relatively large groups were active during the dataset period and some of them were active throughout the period.
\end{abstract}


\section{Introduction}
\label{sec:intro}

While law enforcements agencies and cybercrime researchers are working hard, fake E-commerce scam is still a big threat to Internet users. The number of fake E-commerce sites (hereafter referred to as ``fake EC sites'') that aim to defraud people out of their money or steal their personal information has been increasing, resulting in significant financial damage to society. Many reports indicate that financially-motivated threat actors actively continue scams with fake EC sites~\cite{202401-bayse,202405-SRLabs,jc3_article555, 202411-EclecticIQ}. Additionally, in Japan, the number of reported fake EC sites is on the rise. According to a report from Japan Cybercrime Control Center (JC3)~\cite{jc3_article555}, 47,278 fake EC sites were reported to JC3 in 2023, while 28,818 sites in 2022. Therefore, we need to develop countermeasure to protect Internet users from fake EC sites.

SEO poisoning, or black-hat SEO, using malware installed into compromised websites is one of the major techniques for luring users to fake EC sites, where ``SEO'' stands for ``search engine optimization''. This technique has been observed by multiple security vendors~\cite{lac-lsi-vol2,trendmicro-seo-malware,google-japanese-keyword-hack,sucuri-japanese-keyword-hack}. These malware are installed into compromised websites to intercept web server requests and return arbitrary contents. By doing so, threat actors can send crafted sitemap to search engines, leading them to index generated lure pages as if these were part of the compromised websites. Consequently, search engine users are directed to visit these sites. The malware then intercepts the request handler and redirect user's browser to fake EC sites. 
In this paper, we refer to these compromised websites as ``redirectors'', since they redirect users from search engine results to fake EC sites, and the malware conducts such redirection as ``SEO malware''.
Particularly, the technique which uses Japanese keyword and redirects to Japanese fake EC site is known as ``Japanese Keyword Hack''~\cite{google-japanese-keyword-hack,sucuri-japanese-keyword-hack}.

We previously analyzed the relationship between some malware families used for this strategy~\cite{mshimamura-dsc2024} using information extracted from destined websites.
But the approach lacked completeness because the collection of malware samples was done manually and therefore not all malware families related to black-hat SEO were analyzed. Thus there is an important research question still in a fog; {\it how many threat actor groups conduct fake EC scam in Japan?} To uncover this, we analyze fake EC sites collected in JC3 who monitors cybercrime for a long time. They collected {\numfakesite} fake EC sites trawled from many redirectors over two years, and the collected fake EC sites are filtered by internal confirmation process in JC3, including manual analysis. Thus we expect we can identify more groups than the previous approach. We use data collected in JC3 from {\datadatefrom} to {\datadateto} as a dataset in this study.

To identify groups, we analyzed links between malware family names, fake EC domains, and Matomo~\cite{matomo} servers as key entities in our previous study in~\cite{mshimamura-dsc2024}. But to do the same with the JC3 dataset, we cannot identify SEO malware installed in the redirectors. Thus we analyze links only information extracted from fake EC sites in this study. By analyzing links between entities, we can identify clusters of fake EC sites, whose number approximates the number of threat actor groups. Actually, redirectors can be included for analysis in theory. But we found many redirectors returned only one to three of fake domains during preliminary observation and thus there were difficulty to form groups with redirectors.

Next, based on the groups identified by the link analysis phase, we also conduct a time series analysis, which is described in the later section of this paper. The time series analysis will show trends in group activities and may reveal hidden relationship between groups; for example, if a group become inactive and another group become active instead, we can suspect the latter could be a successor of the former. This hypothesis could be confirmed by characteristics of the two groups. Moreover, if a group operates actively in long time, the group is worth to focus because we can expect the actors behind it will not be inactive soon.

The contributions of this research are threefold:
\begin{enumerate}
    \item We estimated number of threat actor groups conducting fake EC scam with black-hat SEO in Japan, based on a large dataset with in-depth analysis.
    \item Based on the groups identified, we conducted time series analysis to enlighten groups active in the latest situation. 
    \item We also show the effectiveness of our analysis with a case study; multiple groups use new strategy of refund scam on the top of fake EC scam.
\end{enumerate}

The rest of this paper are constructed as follows. 
We describe the background and related works in Section~\ref{sec:related_works} to clarify our research area. 
Section~\ref{sec:dataset} explains the dataset used in this study. 
Section~\ref{sec:link_analysis} presents the result of our link analysis to identify groups. 
Section~\ref{sec:tsa} offers the result of time series analysis based on groups identified in the previous analysis.
In Section~\ref{sec:case_study}, we give a case study of our analysis from warnings in Japanese Consumer Affairs Agency.
We then discuss observations, limitations, and future research directions in Section~\ref{sec:discussion}. 
Finally we summarize this paper in Section~\ref{sec:summary}.

\section{Related works}
\label{sec:related_works}

Kodera,~{\it et al.}~\cite{weko_212864_1} collected and analyzed fake EC sites from search engines using titles common in them and domain names of reported URLs to open blocklists such as URLHaus~\cite{URLHaus} and OpenPhish~\cite{OpenPhish}. Their results show 99.8\% of Japanese fake EC sites, that are actually referred as ``redirectors'' in this paper, redirect users to fake EC sites only when accessing from search engine result. 

Yang,~{\it et al.}~\cite{UsenixSecurity-272118} proposes a technique to detect black-hat SEO for Chinese illegal websites. The target SEO technique of the research is based on a slight defacement in contents of well-known websites, and thus it is different technique from the one focused in this paper.

Zhang,{\it et al.}~\cite{PoisonAmplifier_RAID12} proposes a novel technique to detect compromised websites used for black-hat SEO. This could help our study to collect fake EC sites because it can detect a large number of compromised sites in a week, but we need more consideration as black-hat SEO is a common technique. i.e.) we need to develop an effective filter to detect compromised sites for SEO malware families.


To the best of our knowledge, none of these related studies tried to identify threat actors behind. In this study we try to identify threat actor groups using dataset of redirectors and destined sites. Our goal is to create building blocks to attribute fake EC scam targeting Japan to threat actors. Our previous study~\cite{mshimamura-dsc2024} tried to analyze relationship between fake EC sites and SEO malware families, using Matomo servers and email addresses. The study successfully identified fake EC site groups related to SEO malware families, but as we mentioned in the Introduction, there are lack of completeness because we could not collect ``all'' SEO malware families used in the wild. Instead, we use a large dataset from JC3 in this study to identify threat actors behind Japanese fake EC scams.

\section{Dataset}
\label{sec:dataset}

JC3 dataset includes {\numfakesite} fake EC sites mounted on {\numdomain} domains and {\numemail} email addresses, {\nummatomo} Matomo Server URLs, and {\numid} IDs of 51.la related to them which is summarized in~\tabref{tab:dataset}. The fake EC sites are collected from redirectors by pretending accesses from search engine crawlers and victim users. From the fake EC sites, they extracted email addresses, Matomo Servers and 51.la IDs by regular expression match and a decode program if an email address is encoded by CloudFlare's Web Application Firewall~\cite{cloudflareEmailAddress}.

\begin{table}
\caption{Used dataset from JC3, from May 20, 2022 to Dec. 31, 2024}
\label{tab:dataset}
\begin{center}
\begin{tabular}{l|r}
\hline
\hline 
Type & \# of entities \\
\hline
Fake EC domain & \numdomain \\
Email address & \numemail  \\
Matomo Server & \nummatomo \\
51.la ID & \numid \\
\hline
Total \# of entities for Maltego graph & \numtotalent \\
\hline
\end{tabular}
\end{center}
\end{table}

\section{Link Analysis}
\label{sec:link_analysis}
\subsection{Preliminary group identification}

\begin{figure}[tp]
\begin{center}
\includegraphics[width=0.5\hsize]{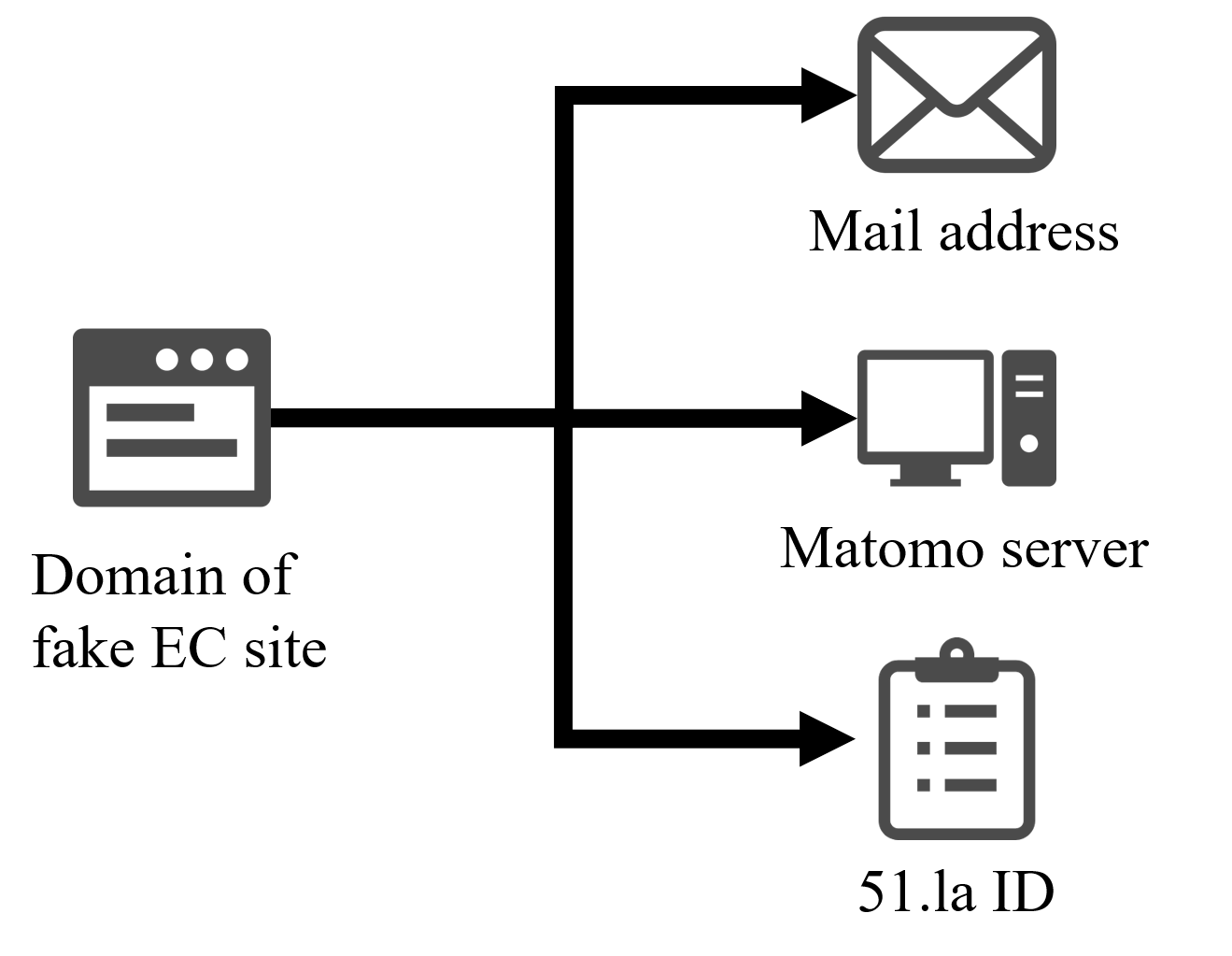}
\end{center}
\caption{Entity connections to create a Maltego graph for link analysis}
\label{fig:entity_map}
\end{figure}

To analyze relationship between redirectors and fake EC sites, we follow our previous approach~\cite{mshimamura-dsc2024}, i.e., we use Maltego~\cite{maltegoHomepage}, a well-known link analysis tool, to visualize relationships as a graph. We define three links to create a graph as followings, depicted in Figure~\ref{fig:entity_map};
\begin{itemize}
\item {\bfseries Link \#1) Domain $\rightarrow$ Mail address} \quad A fake EC site may have multiple mail addresses linked to threat actors. If the same mail address is used in multiple sites, we regard the sites as linked.
\item {\bfseries Link \#2) Domain $\rightarrow$ Matomo server} \quad A fake EC site may have zero or one Matomo server to send information of visitors. If the same Matomo server is used in multiple sites, we regard these sites as related.
\item {\bfseries Link \#3) Domain $\rightarrow$ 51.la ID} \quad 51.la is a well-known access analyzer service often used in fake EC sites. A fake EC site may have zero or one 51.la ID to send information of visitors. If the same ID is used in multiple sites, we regard these sites as related.
\end{itemize}

\begin{figure}[tp]
\includegraphics[width=\linewidth]{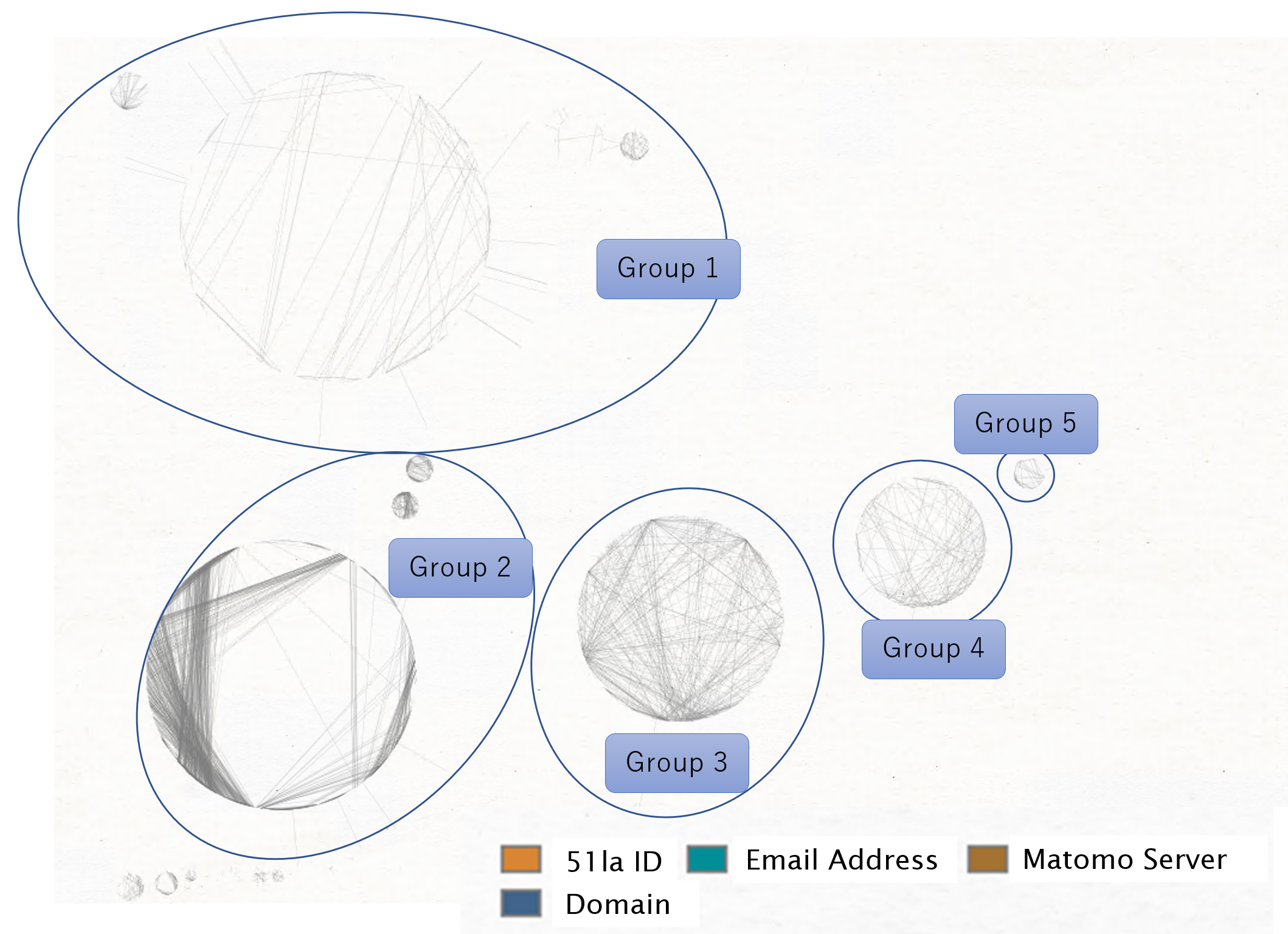}
\caption{A maltego graph shows relationship of fake EC sites}
\label{fig:maltego1}
\end{figure}

\figref{fig:maltego1} is a graph based on the entire dataset, created by Maltego. It shows the entire relationship between fake EC sites, its E-mail address, Matomo servers and redirectors in the dataset.  The graph is shown in ``Circular Layout'' to consider link density. As you can see, there are too many entities and links. Maltego was too heavy to visualize over 100,000 nodes. Thus we created a program to detect groups from the graph file.

The Python code conducts preliminary identification of possible groups from the graph. The code is shown as Listing~\ref{list:grouping}. 
With the code, we detected {\numgroupsall} groups from the dataset. We then filter out groups that have less than {\numthreshold} domains or less than {\numthresholdsites} sites considered as small groups and less priority for analysis. For the rest of this paper focus on the left {\nummaingroup} groups listed in~\tabref{tab:detected_groups}. Note that the group IDs were assigned by the order of domain numbers. G7, G8, and G10 are dropped from focus as they have less than {\numthresholdsites}.
There are cases that multiple subgroups are considered as one group even very weakly connected with just one link due to the algorithm. We will separate them into subgroups by removing such weak links in the next subsection.
Note that it is possible that an actor operates a massive number of sites in small numbers of domains and we mistakenly drop such ``large'' groups, but we surely confirmed dropped 1,110 groups are not the case.

\begin{lstfloat}[!tp]
\begin{lstlisting}[caption=Preliminary grouping algorithm in Python,label=list:grouping,language=Python]
# The entire graph is expressed as "links": a dictionary of links 
# (sourceNode -> [targetNode1, targetNode2, ...]), and
# reverse links (targetNode -> [sourceNode1, sourceNode2, ...])
# for all links in the dataset

def detect_one_group(key, links):
    group = set()
    group.add(key)
    prev_g_len = 0
    # collect all links related to nodes in group
    while prev_g_len < len(group):
        prev_g_len = len(group)
        newg = group.copy()
        for src, dsts in links.items():
            if src in group:
                newg |= dsts
        group = newg
    return group

detected_groups = []
while True:
    try:
        key = list(links.keys())[0]  # a node to begin with
    except: # If unable to select node, finish group detection
        break 
    detected_group = detect_one_group(key, links)
    for k in detected_group:
        del(links[k])  # delete already used links
    detected_groups.append(detected_group)

print(f'Detected {len(detected_groups)} groups')
\end{lstlisting}
\end{lstfloat}

\begin{table}[t]
\caption{Detected groups (G7, G8, and G10 are omitted due to small \# of sites) }

\label{tab:detected_groups}
\begin{center}
\begin{tabular}{l|r|r|r|r|r|l}
\hline
\hline
Group& \# of & \# of  &\# of email & \# of Matomo & \# of & Remarks\\
ID & domains & sites & addresses & servers & 51.la IDs & \\
\hline
G1 & 38,698 & 159,727 & 3,777 & 6 & 3,462 & Has subgroups \\
G2 & 37,665 & 335,787 & 6,517 & 21 & 24 & Has subgroups \\
G3 & 4,897 & 38,888 & 410 & 1 & 0 & Has subgroups,\\
   &       & &     &   &   & Possible relationship with G2 \\
G4 & 1,587 & 7,145 & 83 & 1 & 0 &  \\
G5 & 1,361 & 5,514 & 385 & 3 & 46 &  \\
G6 & 1,343 & 15,433 & 381 & 0 & 105 &  \\
G9 & 352 & 13,080 & 37 & 0 & 39 & \\
G11 & 260 & 2,569 & 8 & 1 & 0 & \\
\hline
Subtotal & 86,163 & 581,430 & 11,598 & 33 & 3,676 & \\
Coverage & 81.84\% & 83.92\% & 86.19\% & 91.67\% & 74.14\% & \\
\hline
\end{tabular}
\end{center}
\end{table}

\begin{table}[t]
\caption{Detected groups and Matomo servers}
\label{tab:groups_matomo}
\begin{center}
\begin{tabular}{l|l}
\hline
\hline
Group ID & Matomo servers (domain only) \\
\hline
G1 & soupn[.]xyz, vxsem[.]xyz, uwmoon[.]xyz, heww[.]xyz \\
G2 & omtage[.]top, ockercsgre[.]top, utermcux[.]top, la51[.]xyz, dvdmoney[.]top,\\
   & axya[.]xyz, phoenixforce[.]xyz, vhuhuzce[.]xyz, alljecknet[.]com, gyfast[.]top, gens2[.]top \\
G3 & la51[.]xyz \\
G4 & onlinea[.]online \\
G5 & piwikcontrol[.]info, piwikfile[.]info, matomotogo[.]site \\
G9 & https[.]or[.]ke, oknice03[.]top\\
\hline
\end{tabular}
\end{center}
\end{table}

\tabref{tab:groups_matomo} describes Matomo server domains used in fake EC sites in the groups. As readers can see, sites in group G3 uses a Matomo server in la51[.]xyz, which is the same Matomo domain used in G2. Thus we believe these two groups are highly related or the same actor group, but we could not confirm relationships from domains, email addresses, and 51.la IDs in them.

\subsection{Detailed identification of subgroups}
As we stated in the previous subsection, there are cases that multiple groups are considered as one group even very weakly connected with just one link due to the algorithm. To split a group into subgroups, we apply the algorithm described in~Listing~\ref{list:grouping} to all cases where the removal of any single entity except for Matomo server from the group graph and extract subgroups that have over {\numthreshold} domains. Note that we except for Matomo servers here because we believe it is less likely that multiple actors share a Matomo server; if multiple groups share Matomo servers, it is not natural there are many Matomo servers working.
We describe the result in~\tabref{tab:detected_subgroups_1}. In the table, group G1 is separated into 28 subgroups named as ``G1-N'' where N is the rank of domain numbers within subgroups of G1. Similarly, group G2 and G3 are also separated into subgroups. 
We applied a two-stage filtering process: first, we extracted subgroups with over {\numthreshold} domains (shown in~\tabref{tab:detected_subgroups_1}), then further filtered to focus on subgroups with over {\numthresholdsites} sites for detailed analysis (shown in~\tabref{tab:detected_subgroups_2}). 
Note that there are cases removing a entity results to three or more groups, and some of the groups are smaller than the threshold, {\numthreshold} domains. We drop such small subgroups if it is the case, rather than merging them with other subgroups. Thus the total size of subgroups may not equal to the size of main group. For example, sum of the numbers of sites for G3-* is 38,859 whereas the number of sites related to G3 is 38,888.

Finally, we identified and focused {\numgroupsfocused} groups shown in~\tabref{tab:detected_subgroups_2}. In the following, we will analyze further for the groups.

\begin{table}[t]
\caption{Number of sites in detected subgroups after first-stage filtering (more than {\numthreshold} domains)}
\label{tab:detected_subgroups_1}
\begin{minipage}[t]{.30\textwidth}
\begin{center}
\begin{tabular}{l|l}
\hline
Subgroup ID & \# of sites \\
\hline
G2-1 & 206,886 \\
G1-1 & 94,627 \\
G2-2 & 61,310 \\
G2-3 & 42,961 \\
G2-4 & 20,382 \\
G1-3 & 17,075 \\
G3-1 & 16,507 \\
G6 & 15,433 \\
G9 & 13,080 \\
G1-4 & 11,248\\
G1-2 & 10,115\\
G3-2 & 9,939\\
G3-4 & 6,297\\
G3-3 & 6,116\\
\hline
\end{tabular}
\end{center}
\end{minipage}
\hfill
\begin{minipage}[t]{.30\textwidth}
\begin{center}
\begin{tabular}{l|l}
\hline
Subgroup ID & \# of sites \\
\hline
G5 & 5,514\\
G11 & 2,569\\
G1-6 & 2,065 \\
\hline
\multicolumn{2}{c}{--- Threshold: ({\numthresholdsites} sites) ---}\\
\hline
G1-14 & 1,901 \\
G1-18 & 1,330 \\
G1-10 & 917 \\
G1-5 & 879 \\
G1-7 & 871 \\
G1-19 & 669 \\
G1-12 & 596 \\
G1-21 & 590 \\
G1-8 & 464 \\
G1-9 & 448 \\
\hline
\end{tabular}
\end{center}
\end{minipage}
\hfill
\begin{minipage}[t]{.30\textwidth}
\begin{center}
\begin{tabular}{l|l}
\hline
Subgroup ID & \# of sites \\
\hline
G1-13 & 373 \\
G1-17 & 363 \\
G1-16 & 356 \\
G1-15 & 348 \\
G1-11 & 340 \\
G1-20 & 311 \\
G1-23 & 286 \\
G1-22 & 285 \\
G1-25 & 243 \\
G1-24 & 228 \\
G1-26 & 211 \\
G1-28 & 210 \\
G1-27 & 201 \\
\hline
\end{tabular}
\end{center}
\end{minipage}
\end{table}

\begin{table}[t]
\caption{Detected subgroups after cutoff}
\label{tab:detected_subgroups_2}
\begin{center}
\begin{tabular}{l|l|l|l|l|l}
\hline
\hline
Group& \# of fake EC & \# of email & \# of Matomo & \# of 51.la & Matomo servers\\
ID & domains & addresses & servers & IDs & (domain only) \\
\hline
G1-1 & 11,822 & 2,345 & 0 & 2,291 & \\
G1-2 & 8,262 & 336 & 0 & 0 & \\
G1-3 & 3,854 & 411 & 3 & 0 & soupn[.]xyz \\
G1-4 & 2,115 & 43 & 3 & 0 & heww[.]xyz, uwmoon[.]xyz, vxsem[.]xyz\\
G1-6 & 696 & 52 & 0 & 0 & \\
G2-1 & 16,314 & 4,403 & 5 & 23 &  la51[.]xyz, gyfast[.]top, omtage[.]top\\
G2-2 & 12,498 & 819 & 5 & 0 &  axya[.]xyz, vhuhuzce[.]xyz\\ 
G2-3 & 5,372 & 469 & 0 & 0 &  \\
G2-4 & 2,986 & 823 & 5 & 1 &  gens2[.]com, utermcux[.]top, \\
 & & & & & alljecknet[.]com, ockercsgre[.]top \\
G3-1 & 1,828 & 114 & 1 & 0 & la51[.]xyz\\
G3-2 & 1,523 & 109 & 0 & 0 &   \\
G3-3 & 853 & 99 & 0 & 0 &   \\
G3-4 & 678 & 86 & 0 & 0 &   \\
G5 & 1,361 & 385 & 3 & 46 &   piwikcontrol[.]info, piwikfile[.]info, \\ 
 & & & & & matomotogo[.]site \\
G6 & 1,343 & 381 & 0 & 105 &   \\
G9 & 352 & 37 & 3 & 39 &   https[.]or[.]ke, oknice03[.]top \\
G11 & 260 & 8 & 0 & 0 &   \\
\hline
\end{tabular}
\end{center}
\end{table}

\section{Time Series Analysis}
\label{sec:tsa}
We conducted time series analysis with the identified groups to track changes. We listed up entities in groups and labeled them. Then we counted how many entities were observed by month.
We split the dataset by months. Because due to the collection period, data of the first month (10 days) and data of the last month (3 days) are not comparable with data of other periods. Thus we remove the periods from time series analysis. 

\begin{figure}[tp]
\includegraphics[width=\textwidth]{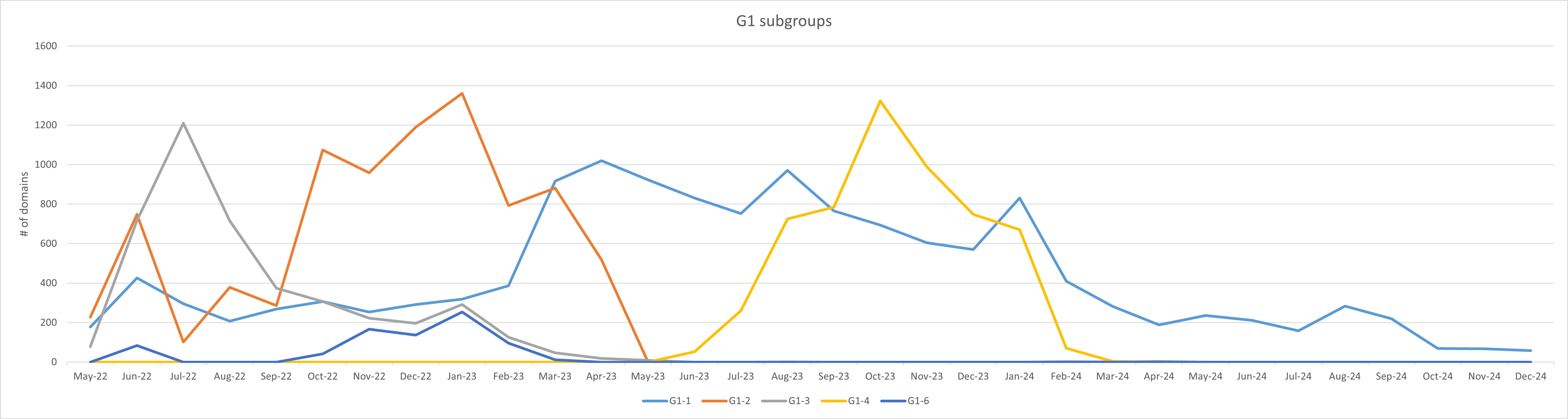}
\caption{Time-chart of fake EC domains in subgroups of G1}
\label{fig:timechart-1}
\end{figure}

\begin{figure}[tp]
\includegraphics[width=\textwidth]{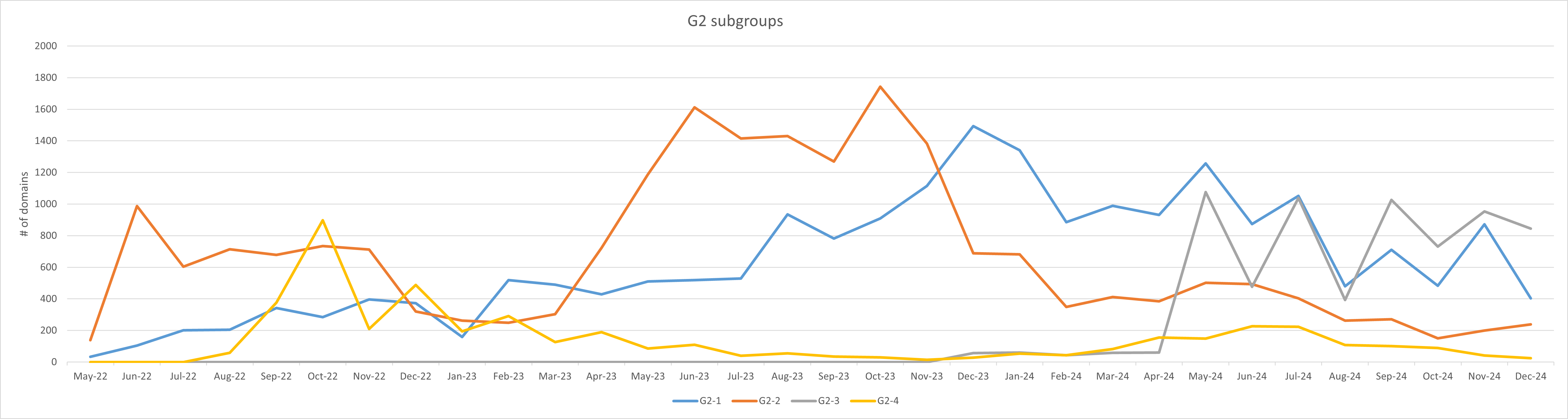}
\caption{Time-chart of fake EC domains in subgroups of G2.}
\label{fig:timechart-2}
\end{figure}

\begin{figure}[tp]
\includegraphics[width=\textwidth]{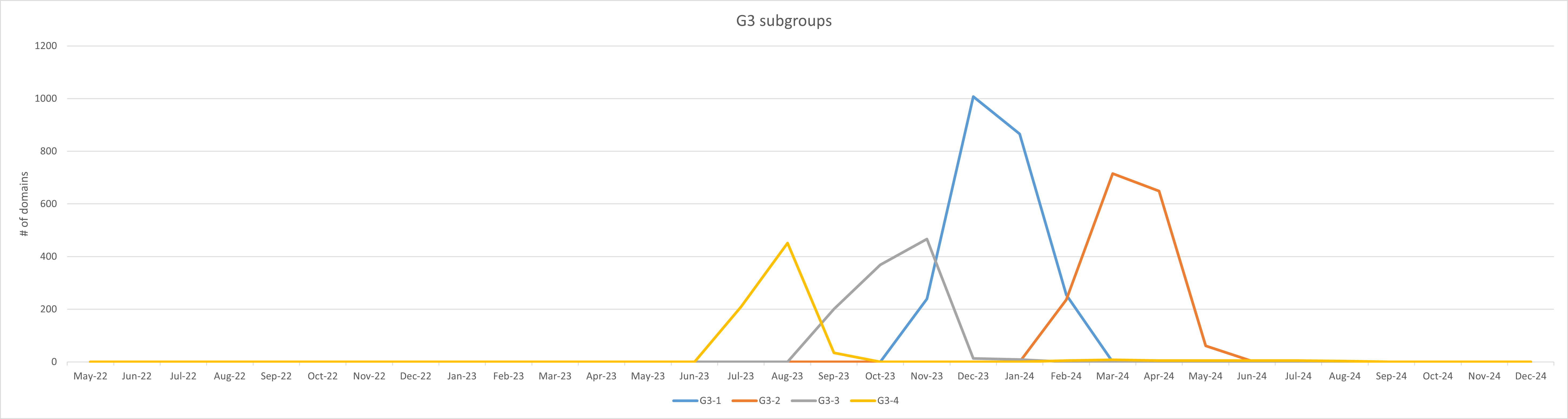}
\caption{Time-chart of fake EC domains in subgroups of G3.}
\label{fig:timechart-3}
\end{figure}

\begin{figure}[tp]
\includegraphics[width=\textwidth]{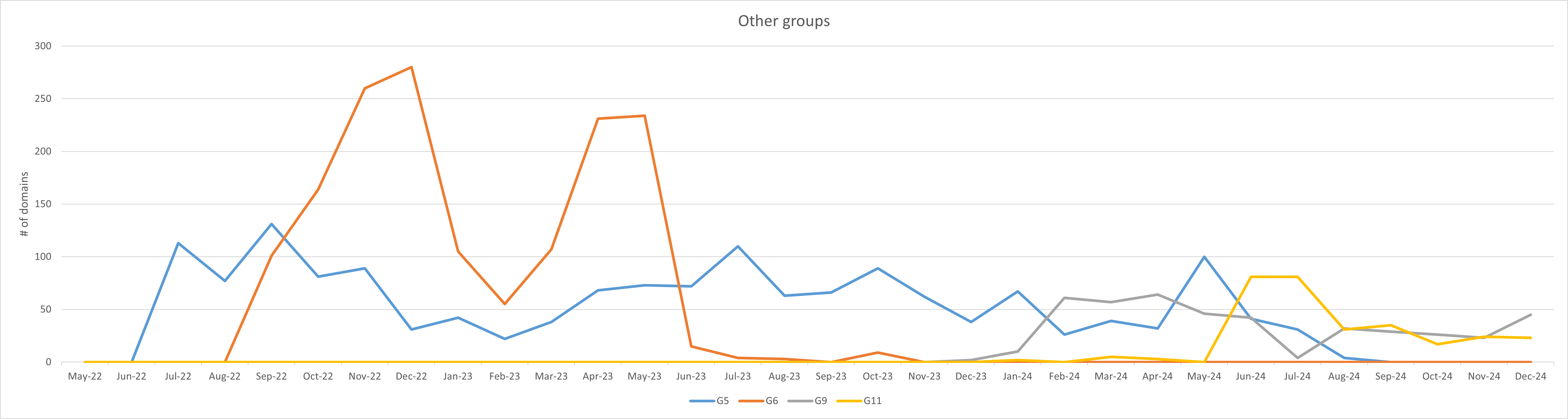}
\caption{Time-chart of fake EC domains in other groups.}
\label{fig:timechart-4}
\end{figure}

The timecharts of numbers of domains in groups are depicted in~\figref{fig:timechart-1},\figref{fig:timechart-2}, \figref{fig:timechart-3}, and \figref{fig:timechart-4} respectively. 
In~\figref{fig:timechart-1}, G1-1 is active throughout the period. On the other hand, G1-2, G1-3, G1-6 shrink their activity in May, 2023. G1-3 appeared in Jun. 2023 to replace them. 

\figref{fig:timechart-2} indicates subgroups of G2 work in relatively long term. On the other hand \figref{fig:timechart-3} illustrate subgroups of G3 operate in a short term, i.e.) less than 4 months.

In \figref{fig:timechart-4}, we can find G5 and G6 already stopped their activities. Notice that group G2-3, G9 and G11 become active in 2024 to the end date of the dataset. We think they are relatively new groups and worth to track.

\section{Case Study: Fake EC sites and Refund Scam}
\label{sec:case_study}

In February 28th, 2025, Japanese Consumer Affairs Agency (CAA) released a public warning for ``refund scam'' in malicious EC sites~\cite{caa_warning_250228}. The scam actor offers a victim for refund to her order with saying the ordered product is out of stock and not able to sell. Then the actor sends message to the victim with a code for a payment application for refunding procedure, but in fact the code is to send money to the actor. The actor then deceive her to push some buttons to send money with the code. The authority reportedly says this scam strategy incurs financial damage of over 680 million yen in about 5,500 cases and publicized 4 sites~\cite{nordot_app_250303}.

We analyzed whether the sites listed in the warning are fake EC sites, and we can attribute to groups if so. The named sites are {\it rdpgk[.]minimumrisk[.]shop}, {\it oggi[.]ayzgyonsale[.]shop}, {\it madrk[.]cnhmxbest[.]shop} and {\it qbague[.]voidnetwork[.]shop}. In the following discussion, we refer them as site ID 1 to 4, respectively. Note that we neutralize all their URLs to prevent readers from unintended access in this paper.

Our analysis reveals that the domains are fake EC sites because they are in the dataset. The detailed analysis result is summarized in~\tabref{tab:refund_scam_groups}. We got exact matches of sites ID 2 and 4 in the dataset, whereas we got no match of sites ID 1 and 3. But we easily identified identified ID 1 as G1-1 because many sites in {\it minimumrisk[.]shop} were linked to it. We also conclude the site ID 3 belong to the group G2-1 without burden because sites that share the same domain were in it.

We also show the column to show whether the domains are related to SEO malware in our previous study~\cite{mshimamura-dsc2024} for readers' information. The sites ID 2 and 3 are related to malware B, and ID 4 is related to malware E. The site ID 1 is not connected to any known SEO malware family but the result infers it could be related to a variant of malware E because sites ID 1 and 4 are linked to the same group G1-1.

As a result of the analysis above above, group G1-1 and G2-1 are related to refund scam cases at least. Surprisingly we find they are the two largest groups. In this case study, the result infers the strategy of refund scam with payment application is possibly shared within multiple fake EC scam groups. 
As we show in this case study, analysis allows us to attribute a scam strategy with fake EC to specific groups once new strategy comes up.
\begin{table}[t]
\caption{Analysis result of URLs related to refund scam publicized from a Japanese CAA warning}
\label{tab:refund_scam_groups}
\begin{tabular}{l|l|l|l|l|l}
\hline
Site & Site URL & Group & Match & Related SEO malware & Related data \\
ID &  & identified  & in dataset & in our previous study~\cite{mshimamura-dsc2024} & first seen \\
\hline
1 & rdpgk[.]minimumrisk[.]shop & G1-1 & Domain & N/A & Sep. 30th, 2024\\
2 & oggi[.]ayzgyonsale[.]shop & G2-1 & Site & Malware B & Oct. 31st, 2024\\
3 & madrk[.]cnhmxbest[.]shop & G2-1 & Domain & Malware B & Sep. 6th, 2024\\
4 & qbague[.]voidnetwork[.]shop & G1-1 & Site & Malware E & Oct. 2nd, 2024  \\
\hline
\end{tabular}
\end{table}

\section{Discussion}
\label{sec:discussion}
\subsection{Necessity for grouping fake EC sites}
The grouping of fake EC sites contributes more detailed analysis, prevention and protection of the threat.
For example, grouping enables us to identify unique characteristics linked to a group.
Actually, in Section~\ref{sec:link_analysis}, we identified some small number of groups use Matomo for analyze visitors, whereas some others use 51.la. More detailed analysis in each group could shed more lights to fake EC actors, but we leave it as our future work. We believe our study in this paper contributes to create a building block to further analysis.

\subsection{Remarks in small groups}
We dropped groups that have less than {\numthreshold} domains or {\numthresholdsites} sites because of limited efforts. But there are some remarkable findings about the dropped groups during study.
\begin{itemize}
\item \textbf{Preference of access analyzer} \quad In~\tabref{tab:detected_groups}, we find dropped groups use only 3 Matomo servers, but 1,282 51.la IDs which is about 26\%. It indicates smaller groups like to use 51.la rather than Matomo. From this result, we get to think small groups may have common tool and manuals that use 51.la.
\end{itemize}

\subsection{Possible countermeasures}
As we pointed out in the above sections, some large group uses Matomo. So the use of suspicious Matomo servers listed in~\tabref{tab:groups_matomo} can be an indicator of fake EC sites. We think replacing Matomo server incurs some burdens on threat actors because relatively small numbers of Matomo server is used while they operate very large numbers of fake EC sites.

On the other hand, unfortunately, there are some difficulties to use 51.la IDs as indicators of fake EC sites. As we show in~\tabref{tab:detected_subgroups_2}, group G1-1 uses 2,291 51.la IDs as of this analysis, and still the number is increasing. So listing all 51.la IDs related to fake EC sites are more difficult than enumerating used Matomo servers. We doubt some actors may use a dashboard app which can integrate multiple 51.la IDs into one view, as 51.la offers API~\cite{51la_api} to export data in realtime.

\subsection{Ethical consideration}
From the viewpoint of research ethics, we thoroughly checked this paper prior to publication.
This study implicitly collects PII (personally identifiable information) including business representatives' names, mail addresses, addresses of fake EC sites, and also information related to PII such as domain names. Most of these information are absolutely fake, or do not allow us to identify threat actors behind them. However, these information could possibly be linked to actual legitimate personnel if a threat actor copied them from other legitimate sites to impersonate, or steal identities from innocent persons. Thus we just show processed data sufficient for discussion in this paper except for URLs in the case study; we did not include any PII and raw data. We also masked URLs, domain names and product names of fake EC sites to avoid conflicts because they might include real trademarks or copyrights.
Note that we believe mentioning original site URLs in Section~\ref{sec:case_study} is no problem because they are confirmed as malicious sites and publicized by the Japanese authority.

On the other hand, we mention some product names like ``Matomo'' and ``51.la'' in this paper, and the Matomo server domains used by threat actors. We think these are important and necessary to include so that readers can reproduce and verify our analysis. The inclusion of these product names in this paper is not likely to damage their reputation.

The redirectors in the dataset are rather victims of website defacement than threat actor related servers. Thus we did not identify any redirectors in this paper. JC3 shares the list of redirectors to law enforcement agencies to reach to administrators of the sites and suggest them to fix, but it is still on the way because the number is huge and often administrators are unreachable, and new redirectors are frequently discovered in their monitoring process. Finally, we must state that JC3 conducts data collection from redirectors in an non-disruptive manner; the request rate is enough low (a request in 5 seconds) not to incur denial-of-services. We believe the access rate is lower than legitimate search engine crawlers and generally acceptable.


\section{Summary}
\label{sec:summary}
In this paper, we analyzed relationship between fake EC sites collected in JC3 to identify threat actor groups. Based on the fake EC site dataset collected from {\datadatefrom} to {\datadateto}, we identified {\numgroupsall} possible groups and then filtered to {\numgroupsfocused} groups based on their size for more detailed analysis. Then we conducted time series analysis for the {\numgroupsfocused} groups to analyze how long they are active. The result of time series analysis shows that some groups were active throughout the period of the dataset and some other groups were vanished in a short term. Also we give a case study using shared URLs from Japanese Consumer Affairs Agency and shows new refund scam strategy are used in multiple fake EC groups.

We hope the groups identified based on this analysis can help law enforcement to track their activities.


\bibliographystyle{IEEEtran}
\bibliography{references}  






\end{CJK}

\end{document}